# Highly Nonlinear and Low Confinement Loss Photonic Crystal Fiber Using GaP Slot Core

Md Borhan Mia[1*], Animesh Bala Ani[1], Kanan Roy Chowdhury[2], and Mohammad Faisal[1]
[1]Bangladesh University of Engineering and Technology, Dhaka, Bangladesh
[2]Chittagong University of Engineering and Technology, Chittagong, Bangladesh
*mborhanm1110@gmail.com

*Abstract*—This paper presents a triangular lattice photonic crystal fiber (PCF) with very high nonlinear coefficient. Finite element method is used to scrutinize different optical characteristics of proposed highly nonlinear PCF (HNL-PCF). The HNL-PCF exhibits a high nonlinearity up to $10 \times 10^4$ $W^{-1}km^{-1}$ over the wavelength band of 1500 nm to 1700 nm. Moreover, proposed fiber shows a very low confinement loss of $10^{-3}$ dB/km and chromatic dispersion of −9000 ps/(nm-km) at 1550 nm wavelength. Furthermore, dispersion slope, effective area are also analyzed thoroughly. The proposed fiber will be a suitable candidate for sensing applications, dispersion compensation, nonlinear signal processing and supercontinuum generation.

## I. INTRODUCTION

Photonic crystal fibers or microstructure holey fibers exhibit diversified properties which furnish some new applications such as supercontinuum generation, fiber sensors and ability to uphold high polarization, broadband dispersion controlling etc. Nonlinearity is one of the indispensable belongings of photonic crystal fibers for many useful applications including optical switching, optical regeneration, supercontinuum generation, optical parameter magnification, and optical wavelength transformation [1], [2]. Compared to standard single mode fibers (SMFs), PCFs have many tunable properties, for example; air hole diameter, pitch, cladding, background material, doped core etc. These flexibilities provide better control over nonlinearity, dispersion slope, birefringence, splice loss and confinement loss etc. These are only possible in PCF which are unachievable in SMFs. PCFs are classified into two groups, index guiding and photonic band gap PCFs. In these two types of PCFs, high refractive index difference is upheld in the middle of core and cladding.

To attain high nonlinearity researchers have studied the behavior of PCFs by using nanostructure with large refractive index in the core section. Using pure silica core, nonlinear coefficient is only about 100 $W^{-1}km^{-1}$ due to very small nonlinear refractive index of silica, nominally $29.6 \times 10^{-21}$ $m^2/W$. Therefore, higher nonlinear refractive index materials are used in the core to improve nonlinearity. Recently, Liao et al. [3] suggested a PCF of high nonlinearity using nano scale slot core. The PCF exhibits a very high nonlinearity up to $3.5739 \times 10^4$ $W^{-1}km^{-1}$. However, confinement loss issue is ignored. Huang et al. proposed a slot coiled silicon PCF having a high nonlinear coefficient up to 1068 $W^{-1}m^{-1}$ [4]. Li and Zhao used nano wires of gold in core and attained boosted polarization reliant coupling and transmission [5]. Liao et al. [6] suggested a spiral PCF of high nonlinearity exhibiting nonlinear coefficient of 226 $W^{-1}m^{-1}$ at the communication band. Amin et al. [7] proposed a spiral high nonlinear photonic crystal fiber using GaP strips in the core. The fiber shows a high nonlinearity of $10^4$ $W^{-1}km^{-1}$. Nevertheless, fiber exhibits confinement loss of $10^3$ and $10^{-10}$ dB/km for x and y polarization modes, respectively at 1550 nm wavelength. To fabricate fibers having large nonlinearity with nanoscale slot core, recently article [8] is published.

In our article, a triangular lattice photonic crystal fiber is being suggested which shows a very high nonlinearity up to $8 \times 10^4$ $W^{-1}km^{-1}$ at wavelength of 1550 nm. It exhibits nonlinearity of $10 \times 10^4$ to $2 \times 10^4$ $W^{-1}km^{-1}$ in wavelength array of 1550 nm to 1770 nm. Additionally, it shows a very minute confinement loss of $10^{-3}$ dB/km at communication band. To our knowledge this is the best result compared to recently published articles. Moreover, HNL-PCF shows a very high dispersion up to −9000 ps/(nm-km) at 1550 nm wavelength. Therefore, suggested fiber can be useful supercontinuum generation, optical parameter amplification and broadband dispersion compensation.

## II. FIBER DESIGN

The geometric view of the proposed HNL-PCF with magnified sight of slot core is demonstrated in Fig. 1. The design is kept as simple as possible. The cladding region comprises of six air hole rings. The air hole diameter in cladding region, d = 0.58 μm with the pitch value of Λ = 0.75 μm. Air filling fraction in cladding is d/Λ = 0.77 which is fabrication feasible. Two alike rectangular strips of GaP are familiarized in core with a space of $L_s$ = 0.12 μm. The length and width of the strips are $d_x$ = 0.58 μm and $d_y$ = 0.096 μm, respectively. The amplified view of the core in Fig. 1(a) exhibits $d_x$, $d_y$ and $L_s$. The background material is silica ($SiO_2$) which is industrially accessible.

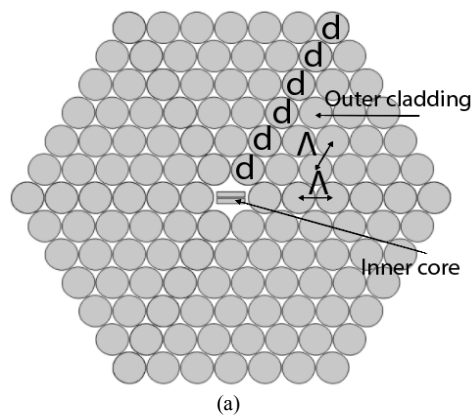

(a)

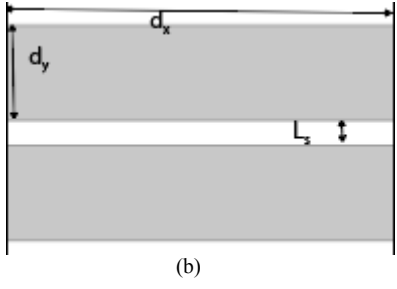

(b)

Fig. 1 (a) Geometric view of the suggested HNL- PCF (b) magnified view of core.

### III. SIMULATION AND RESULT

Finite Element Method (FEM) is used to study different optical and electrical characteristics of the proposed design. COMSOL MULTIPHYSICS 5.0 is used to simulate the full design. Modal analysis has been studied by solving the Eigen value problems drawn from Maxwell's equations. A perfectly matched layer (PML) is placed at the outmost ring to avert reflection [9]. Since refractive indices depend on wavelength, Sellmeier's constants for silica (SiO$_2$) and GaP are directly employed in simulation to improve the accuracy. Once modal refractive index, $\eta_{eff}$ is obtained, other parameters like chromatic dispersion $D(\lambda)$, nonlinear coefficient $\gamma$, confinement loss $L_c$ and effective area $A_{eff}$ be able to be tailored from their equations [11]–[14].

$$\gamma = \frac{2\pi n_2}{\lambda A_{eff}} \quad (1)$$

$$A_{eff} = \frac{\left(\iint |E|^2 \, dxdy\right)^2}{\iint |E|^4 \, dxdy} \quad (2)$$

$$L_c = \frac{20 \times 10^6}{\ln(10)} k_0 Im[\eta_{eff}] \quad (3)$$

$$D(\lambda) = -\frac{\lambda}{c} \frac{d^2 \, Re[\eta_{eff}]}{d\lambda^2} \quad (4)$$

Where, $Re[\eta_{eff}]$ and $Im[\eta_{eff}]$ are the real and imaginary part of the effective refractive index, respectively; $c$ is the velocity of light in void, $\lambda$ is the operating wavelength, $E$ is the field vector, $k_0 = \frac{2\pi}{\lambda}$, is the wave number in vacuum; $n_2$ is the Kerr constant. For GaP, the $n_2 = 6.63 \times 10^{-18}$ m$^2$W$^{-1}$ [10]. Three termed Sellmeier formula for silica (SiO$_2$) and GaP are directly included in simulation.

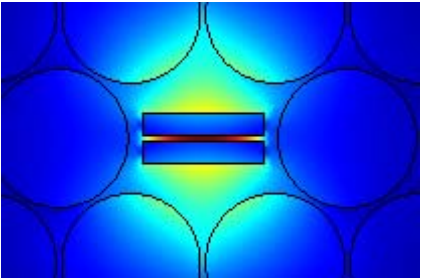

Fig. 2 Poynting vector profile of the proposed HNL-PCF.

Fundamental mode field distribution of the proposed fiber with parameters Λ = 0.75 μm, d = 0.58 μm, d$_x$ = 0.5 μm, d$_y$ = 0.096 μm and L$_s$ = 0.12 μm is depicted in Fig. 2. It is seeming that fundamental mode field is well restrained in slot section of GaP meaning lesser effective mode be able to be achieved from the proposed fiber. Dependence of the effective mode on wavelength of the proposed fiber is explained in Fig. 3. From Fig 3, it is understood that effective area at 1550 nm wavelength is 0.3 μm$^2$, which is very small. This small area of the effective mode gives rise of huge nonlinearity.

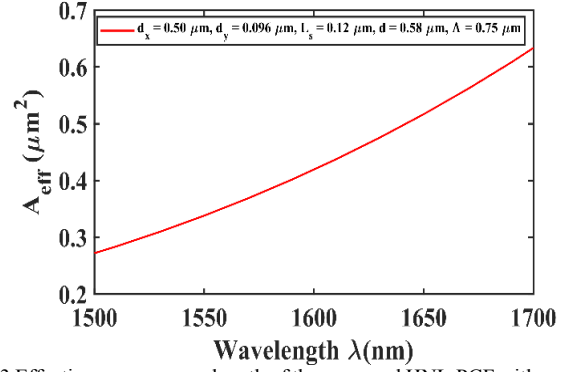

Fig. 3 Effective area vs. wavelength of the proposed HNL-PCF with parameters Λ = 0.75 μm, d = 0.58 μm, d$_x$ = 0.5 μm, d$_y$ = 0.096 μm and L$_s$ = 0.12 μm.

Fig. 4 (a) shows the effects of variation in length of GaP strip on nonlinearity. It is apparent that increasing strip length d$_x$ results in decreasing nonlinearity from 1500 nm to 1700 nm. At 1500 nm, nonlinearity is up to 10 × 10$^4$ W$^{-1}$km$^{-1}$ whereas it would be up to 4 × 10$^4$ W$^{-1}$km$^{-1}$ at 1700 nm wavelength. Almost alike outcomes are deduced for increasing strip width d$_y$ in Fig. 4 (b). For lower strip width, HNL-PCF has higher nonlinearity. This is because, by decreasing the strip length or width improves the light confinement in HNL-PCF which reduces the effective area and enhances the nonlinear coefficient.

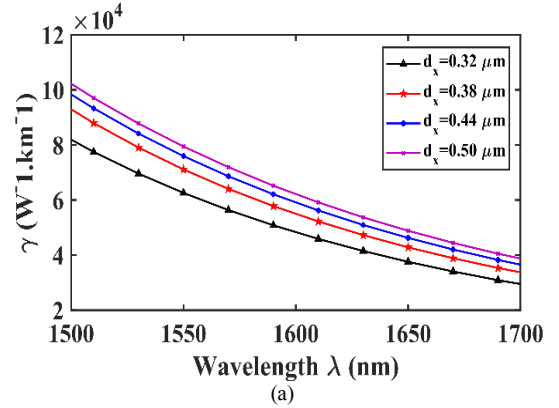

(a)

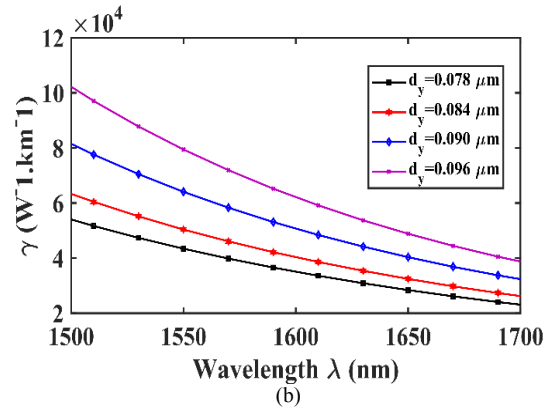

(b)

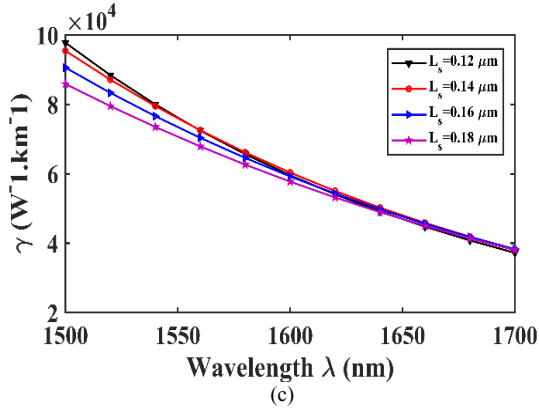

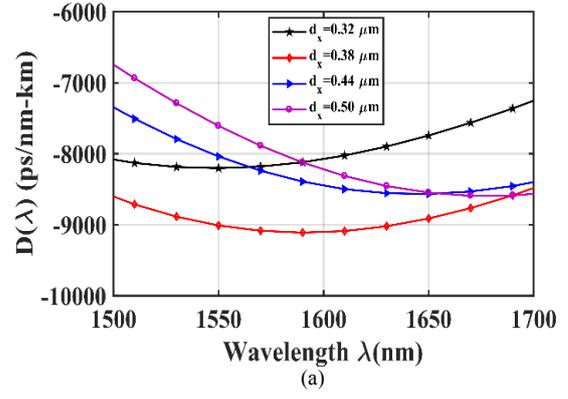

Fig. 4 Nonlinear coefficient vs. wavelength for variation of (a) strip length $d_x$, with $d_y = 0.096$ μm, $L_s = 0.12$ μm, $Λ = 0.75$ μm and $d = 0.58$ μm; (b) strip width $d_y$, with $d_x = 0.50$ μm, $L_s = 0.12$ μm, $Λ = 0.75$ μm and $d = 0.58$ μm and (c) slot width $L_s$, with $d_x = 0.50$ μm, $d_y = 0.096$ μm, $L_s = 0.12$ μm, $Λ = 0.75$ μm and $d = 0.58$ μm.

Fig. 4 (c) exhibits the effects of changing slot width on nonlinearity. It is clear that, increasing $L_s$ reduces nonlinearity. At 1550 nm wavelength nonlinearity is $9.8 × 10^4$ W$^{-1}$km$^{-1}$ when $L_s = 0.12$ μm. However when $L_s = 0.18$, nonlinear coefficient reduces to $8.4 × 10^4$ W$^{-1}$km$^{-1}$. Since the slot width effective area increases, results in subsequent decrease in nonlinearity.

Chromatic dispersion of the suggested HNL-PCF is demonstrated in Fig 5(a), 5(b) and 5(c) by altering the parameters of $d_x$, $d_y$ and $L_s$, respectively. In Fig 5(a), chromatic dispersion decreases with the increment of strip length $d_x$. At the communication band dispersion reaches up to −9000 ps/(nm-km). In Fig 5(b), strip width $d_y$ is altered to see the effect of it. With the increment of strip width $d_y$, chromatic dispersion also increases and reaches up to −7500 ps/(nm-km) at 1550 nm wavelength. The effect of slot width $L_s$ is explored in Fig. 5(c). From Fig. 5(c), it is clear that with the maximum slot width, we obtain the less dispersion. When the slot width $L_s = 0.12$ μm, we acquire maximum dispersion of −8000 ps/(nm-km) at communication band.

Confinement loss of the suggested HNL-PCF is depicted in Fig 6. In Fig 6(a), strip length of GaP is varied keeping other parameters constant. With the increment of strip length $d_x$, confinement loss decreases. Similar effect is shown for the increment of strip width $d_y$. In both, case confinement loss can be as low as $10^{-3}$ dB/km. However, slot width has opposite effect. With the slot width $L_s$, confinement loss increases.

In simulation, we varied the slot width $L_s$, strip length $d_x$ and strip width $d_y$ keeping other parameters unaltered; i.e., pitch $Λ = 0.75$ μm and air holes have diameter of $d = 0.58$ μm due to core length and width provided major effect on nonlinearity than cladding. Moreover, chromatic dispersion of the suggested HNL-PCF mainly depends on the $L_x$, $d_x$ and $d_y$ and these are justified by varying the core diameter which has less effect than strip width, length and slot width.

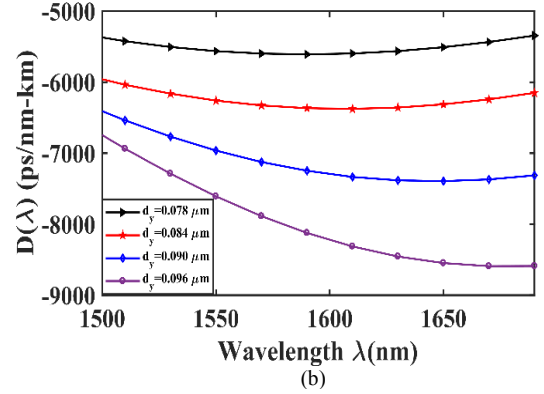

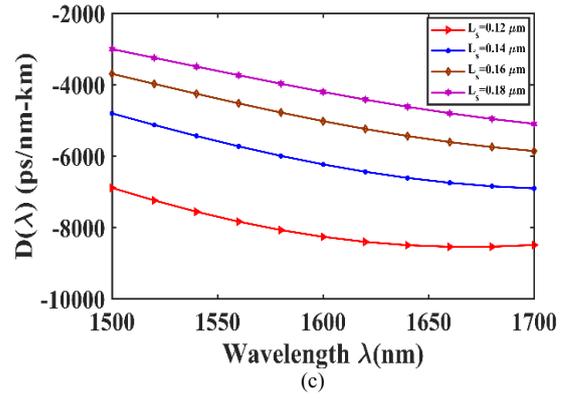

Fig. 5 Effects on chromatic dispersion depending on wavelength for variation of (a) strip span $d_x$, with $d_y = 0.096$ μm, $L_s = 0.12$ μm, $Λ = 0.75$ μm and $d = 0.58$ μm; (b) strip width $d_y$, with $d_x = 0.50$ μm, $L_s = 0.12$ μm, $Λ = 0.75$ μm and $d = 0.58$ μm and (c) slot width $L_s$, with $d_x = 0.50$ μm, $d_y = 0.096$ μm, $L_s = 0.12$ μm, $Λ = 0.75$ μm and $d = 0.58$ μm.

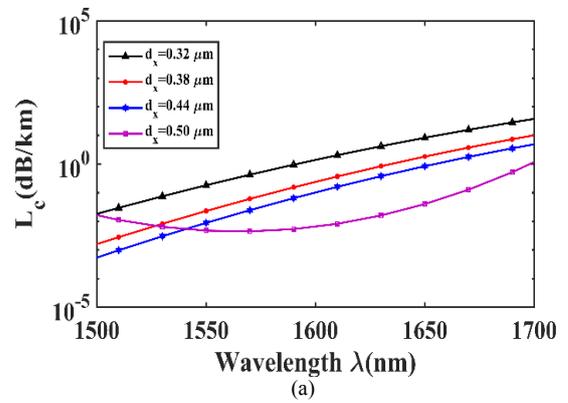

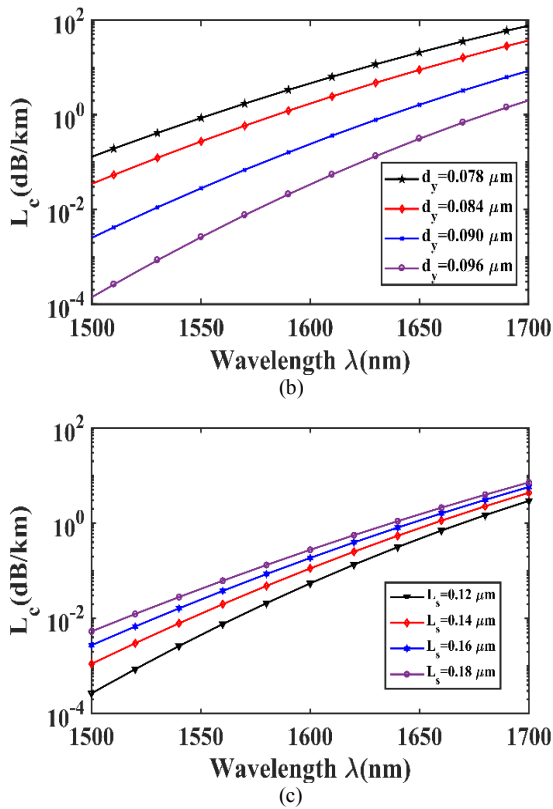

Fig. 6 Confinement loss vs. wavelength for variation of (a) strip span $d_x$, with $d_y$ = 0.096 μm, $L_s$ = 0.12 μm, Λ = 0.75 μm and d = 0.58 μm; (b) strip width $d_y$, with $d_x$ = 0.50 μm, $L_s$ = 0.12 μm, Λ = 0.75 μm and d = 0.58 μm and (c) slot width $L_s$, with $d_x$ = 0.50 μm, $d_y$ = 0.096 μm, $L_s$ = 0.12 μm, Λ = 0.75 μm and d = 0.58 μm.

## IV. Conclusion

In our paper, we numerically examined different optical properties of HNL-PCF using FEM and COMSOL MULTIPHYSICS as simulator. The proposed HNL-PCF shows a very high nonlinearity up to $10 \times 10^4$ W$^{-1}$km$^{-1}$ and a very low confinement loss of $10^{-3}$ dB/km at 1550 nm wavelength. The fiber can have potential application in supercontinuum generation, optical parameter amplification and broadband dispersion compensation. To our best knowledge, nonlinearity that HNL-PCF shows is the highest compared to recently published articles.